\documentstyle[prl,aps]{revtex}
\begin{document}
\title{Point-contact spectroscopy on URu$_2$Si$_2$}
\author{J.G. Rodrigo$^a$, F. Guinea$^b$, S. Vieira$^a$, and F.G. Aliev$^c$.}
\address{Instituto Universitario de Ciencia de Materiales ``Nicol\'as
Cabrera''\\$^a$%
Laboratorio de Bajas Temperaturas. Departamento de F\'\i sica de la Materia
Condensada.\\Universidad Aut\'onoma de Madrid. 28049 Madrid, Spain.\\$^b$%
Instituto de Ciencia de Materiales, CSIC. 28049 Madrid, Spain.\\$^c$%
Laboratorium voor Vaste Stoffysica en Magnetisme. Katholeike Universiteit
Leuven.\\Celestijnenlaan 200D, B3001 Leuven, Belgium.}
\date{\today }
\maketitle

\begin{abstract}
Tunnel and point contact experiments have been made in a URu$_2$Si$_2$
single crystal along the {\em c}-axis. The experiments were performed
changing temperature and contact size in a low temperature scanning
tunneling microscope. A resonance develops at the Fermi level at $T\sim 60$
K. This resonance splits and becomes asymmetric when the 17.5 K phase
transition is crossed. These results are consistent with the existence of
Kondo like bound states of the U$^{4+}$ ionic configurations and the
conduction electrons. Below the transition, these configurations are split
by the development of quadrupolar ordering. The peak separation can be
interpreted as a direct measurement of the order parameter. Measurements on
a policrystalline UAu$_2$Si$_2$ sample are also reported, with a comparative
study of the behavior of both materials.
\end{abstract}


A large effort has been made in the last years to understand the low
temperature behaviour of uranium based ternary intermetallic alloys.
Magnetic and superconducting phase transitions have been found on UT$_2$Si$%
_2 $ compounds (T refers to a transition element), which have been the
subject of intensive experimental and theoretical studies. A comprehensive
explanation of the behavior of these materials is, however, still lacking.
One the most studied compound of this family is URu$_2$Si$_2$. Among other
remarkable features, this compound exhibits antiferomagnetism below 17.5K
and superconductivity below 1K. The nature of the pairing, and the origin of
the transition at 17.5K are not completely understood. Different experiments
including specific heat, thermal expansion, magnetic susceptibility,
resistivity, inelastic neutron scattering, and tunnel and point contact
spectroscopies, have been made, with special emphasis in the search of an
explanation of the 17.5 K transition. The fact that the measured magnetic
moment per uranium atom in the magnetically ordered phase is 0.04 $\mu _B$,
is in contrast with the measured moments per uranium atom, 1.6-3.0 $\mu _B$,
for the other UT$_2$Si$_2$ alloys on which neutron scattering experiments
have been performed.

As pointed in ref.\cite{mfield}, it is difficult to identify such a small
magnetic moment with the order parameter of an antiferromagnetic phase
transition with N\`eel temperature of 17.5K, if we take account the large
anomalies observed in several quantities at this temperature. Based on these
experimental facts, Ramirez {\em et.al.}\cite{Ramirez} proposed several
years ago that the order parameter is not the tiny magnetic moment\cite
{neutron}. This is a secondary effect originated because the fundamental
order parameter breaks time-reversal symmetry, which allows a linear
coupling to the staggered magnetic order, inducing the measured magnetic
moment (see, however\cite{Walker}, for an alternative explanation).
Following this idea, a model based on quadropolar ordering of localized {\em %
f} electrons has been developed in ref.\cite{mfield}, which reproduces
semiquantitatively the magnetic properties of URu$_2$Si$_2$. The model
assumes an effective coupling between the U$^{4+}$ configurations in
different U atoms, which is presumably induced by the conduction electrons.
The level ordering which best fits the experimental results has three
singlets at low energies, while the doublets lie at higher energies. This
model has been generalized in\cite{spin}, by also including the magnetic
configurations of the U$^{3+}$ and U$^{5+}$ charge states.

In the following, we will try to gain information on the U ionic
configurations, and their coupling to the conduction electrons, by means of
tunnel and point contact spectroscopy. Electron tunneling and point-contact
spectroscopies have been useful tools to study electronic properties of
several UT$_2$Si$_2$. Several experiments have been made during the last
years\cite{Aliev,pc-exp,stm-exp,Escudero}, although a consistent
interpretation in terms of the hybridization of the U configurations and the
conduction electrons is still lacking.

In previous works we have used a scanning tunneling microscope (STM), which
we can cover temperatures ranging from 2 K up to 300 K operating inside a $%
^4 $He cryostat\cite{tesis}, to study the conductance behaviour of
nanoscopic size junctions between different materials. The electrodes are
displaced, in a controlled way, from a vacuum tunneling regime to point
contact\cite{expPband}. Similar procedures were followed in the present
experiments. The URu$_2$Si$_2$ samples were thin slices cleaved
perpendicular to the c-axis from a high quality single crystal, which has
been characterized by measuring magnetic susceptibility\cite{previos} and
heat capacity. The specific heat showed a sharp peak (width less than 0.5 K)
near 17 K. The sample also shows a superconducting transition at T$_c\simeq
1.2$ K, with a transition width of $\sim 0.1$ K. Noble metals, platinum and
gold, were used as tip counterelectrodes. The experiments were performed by
changing the temperature, between 4.2 K and 40 K, and changing the area of
the contact. We compare these results with the ones obtained in an UAu$_2$Si$%
_2$ policrystalline sample, which presents ferromagnetic ordering near 17 K,
and with junctions formed with two URu$_2$Si$_2$ electrodes.

At a selected spot on the sample current vs. voltage, {\it I(V),} and
current vs. tip-to-sample distance, {\it I(d)}, curves can be taken.
Typically, these curves are taken in 100 ms with 1024 data points.
Differential conductance curves, $dI/dV(V)$, are obtained by numerical
derivation of the {\it I(V)} data. Varying the z-piezo voltage we can cover
vacuum tunneling and clean contact regimes with resistance ranging from
several M$\Omega $ down to a few ohms.

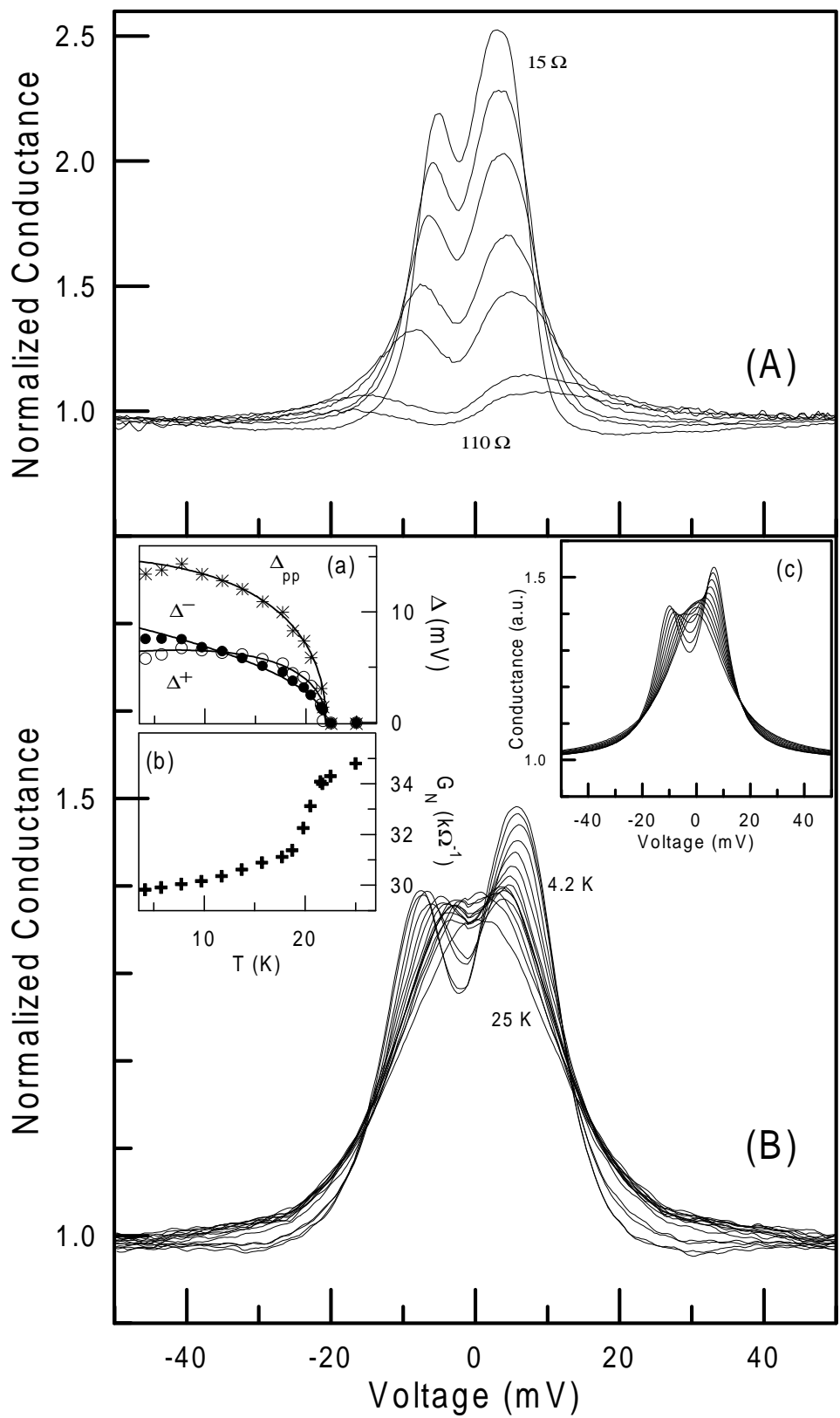
\begin{figure}[]
\caption{A) Normalized conductance spectra taken at 4.2 K with different contact
resistance for Pt - URu$_2$Si$_2$ contacts.
$R_N$ values are: 15, 20, 25, 40, 50, 105 and 110 $\Omega $ (from top to
bottom).
B) Normalized conductance spectra of a Pt - URu$_2$Si$_2$ contact,
with temperature varied from 4.2 K to 25 K. Insets: (a) evolution of the peaks
positions
with temperature. Lines are fits to the dependences
proposed in the text; (b) variation of the conductance with temperature;
(c) simulation of the evolution of the URu$_2$Si$_2$ conductance
with temperature using two lorentzian functions. }
\label{fig1}
\end{figure}

Every experimental round starts at 4.2 K with topographical and
spectroscopical explorations. {\it I(d)} curves are taken to check the
cleanliness of the contact\cite{limpio}. Then, series of {\it I(V)} curves
are taken for different junction resistances. The main feature in the
conductance curves is a central hump at zero bias, which presents two
asymmetric peaks respect to bias voltage (Fig.\ref{fig1} (A) ).

Once we have reached the contact regime, we study the evolution of different
features in the {\it dI/dV(V) }curves as a function of temperature in
repeated cycles up to 40 K. These features are the position of conductance
peaks, $\Delta ^{+}$ and $\Delta ^{-}$, its separation $\Delta _{pp}$, and
the normal conductance of the contact, $G_N$, (i.e. the value of {\it dI/dV }%
at high voltage, $\sim 100$ mV, because no heating effects are detected).

In Fig. \ref{fig1}(B) we present a series of conductance curves
corresponding to one thermal cycle. We observe that, as the temperature
increases from 4.2 K the peaks move to zero voltage, they evolve into a
single hump at about 20 K, then this hump broadens and decreases its height
until it almost disappears at higher temperatures. This evolution can be
extrapolated, and we find the onset of this resonance at Fermi level at $%
T\simeq 60$ K.

The insets in Fig. \ref{fig1}(B) show the evolution of the parameters
mentioned above. In inset (a) we present the evolution of the conductance
peaks: for low temperatures $\Delta ^{+}$ is smaller than $\Delta ^{-}$ ,
which can be extrapolated to $\Delta _0^{+}\simeq 6$ meV and $\Delta
_0^{-}\simeq 8$ meV at $T=0$ K, and at higher temperatures they decrease
uniformly, having $\Delta ^{\pm }=0$ at $T_N\simeq 22$ K. From these data,
we consider $T_N\simeq 22$ K as the transition temperature to the
antiferromagnetic state in our experiment.

Inset (b) shows the evolution of $G_N$: an increase of $G_N$ (and therefore
an increase of the electronic density of states) respect to the value at low
temperature takes place when $T>T_N$, as has been previously stated by other
authors from specific heat measurements \cite{inc-DOS}. We observe this
effect both in series taken in constant current mode (there is a clear
change in the feedback signal when $T_N$ is crossed), and in series taken
with fixed voltage applied to the piezo.

There is one question concerning the {\em high} value that we obtain for $%
T_N $, compared with the value commonly found in the literature ($T_N\simeq
17$ K). In our experiments we are doing point-contact spectroscopy using a
STM, with reasonably clean contacts, whose radii are about 50 \AA . In these
conditions, in similar STM experiments \cite{fuerzas}, it has been shown
that the pressure at the contact is of the order of 1 GPa ($\sim $10 kbar).
These values can be reached because the fact that the STM tip is small.
Resistivity measurements in URu$_2$Si$_2$ have shown an increase of $T_N$
with pressure \cite{ro-presion}. The behaviour of $T_N$ vs $P$ is in good
quantitative agreement with our observation of $T_N\simeq 22$ K for $P\sim
10 $ kbar. This effect can also be noticed in other spectroscopic works on
URu$_2$Si$_2$. In point-contact measurements using an STM\cite{stm-exp}, the
double peak in {\it dI/dV }develops at temperatures similar to those in our
experiment. When standard point-contact is used\cite{pc-exp,Escudero},
involving larger contact areas than in STM experiments, and thus less
pressure at the contact, the double peak develops at temperatures closer to $%
T_N(P=1\,$bar$)\simeq 17$ K.

If we pay attention to the evolution of {\it dI/dV(V) }series at fixed
temperature and varying the contact resistance (Fig.\ref{fig1}(A)), we
observe that the double peaked central structure increases its height as we
attain higher conductance values (i.e., larger size of the contact). This is
a common situation when inelastic scattering processes are involved in
point-contact spectroscopy measurements\cite{pc-clasico}. In this case, the
larger the ratio $a/l$ (with $a$ radius of the contact, and $l$ inelastic
electron mean free path), the larger will be the information about inelastic
processes that appears in the conductance spectra. This is usually observed
in measurements of the phonon structure in normal metals.

The double peak structure found experimentally can be well approximated by
two (or three, see below) lorentzians whose separation is proportional to $%
\sqrt{T_N-T}$, and whose average position deviates from the Fermi level as $%
T_N-T$. A simulation of the evolution of the conductance with temperature is
shown in inset (c) of Fig.\ref{fig1}. We have used two states (lorentzians)
with the same initial width (12 mV) that increases like $T^2$, and with
heights ratio of 1.33. Temperature varies from 4 K to 22 K.

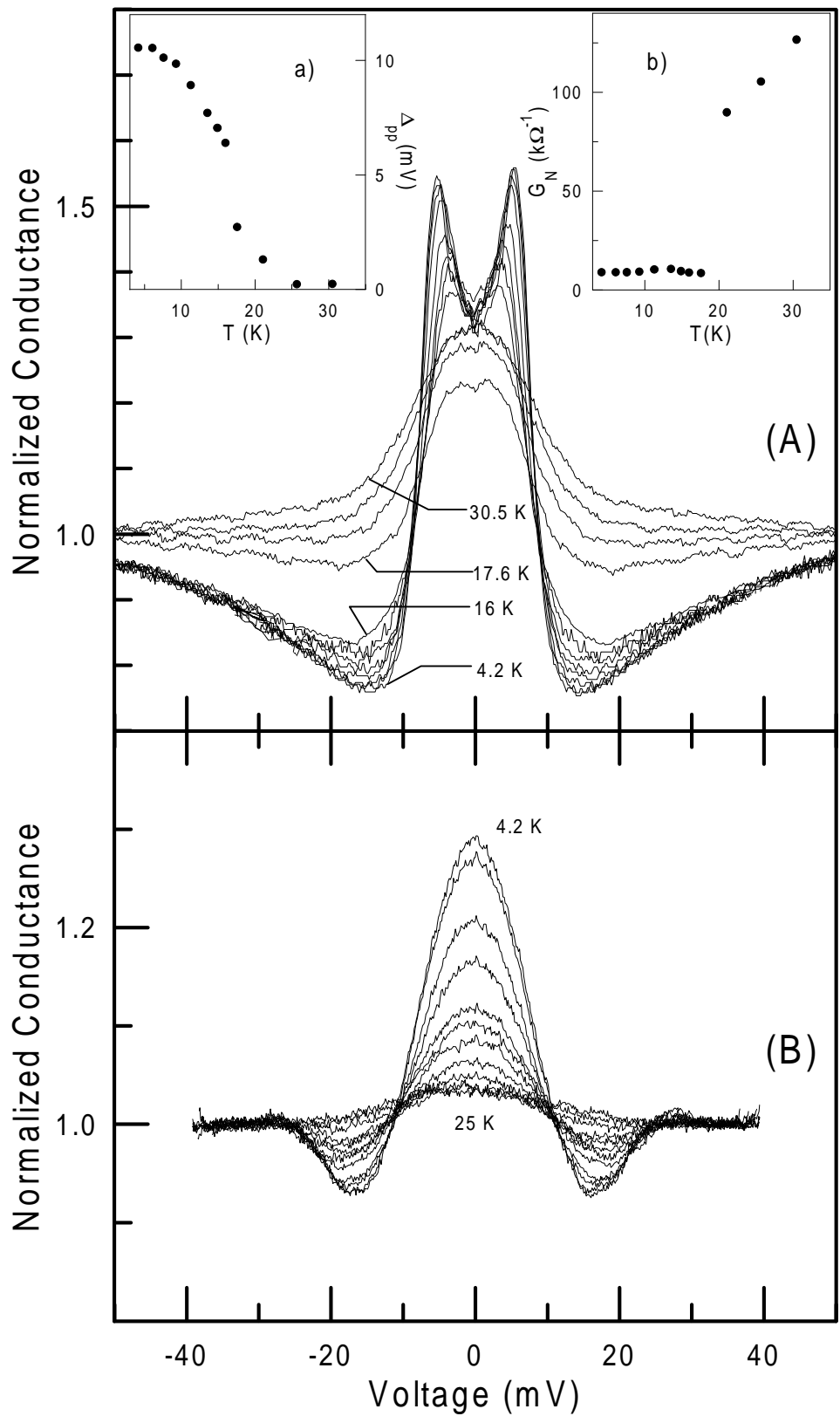
\begin{figure}[]
\caption{A) Normalized conductance spectra obtained in the range
4.2 K $\leq $ T $\leq $ 30.5 K for a URu$_2$Si$_2$ - URu$_2$Si$_2$ contact.
Insets: evolution of peaks separation (a) and normal conductance (b)
with temperature.
B)  Normalized conductance spectra obtained
in the temperature range 4.2 K $\leq $ T $\leq $ 25 K for a
Pt - UAu$_2$Si$_2$ contact with $R_N\sim 50$ $\Omega $.}
\label{fig2}
\end{figure}

In Fig. \ref{fig2}(A) we present the evolution of {\it dI/dV }vs. $T$ for
symmetric URu$_2$Si$_2$ contacts\cite{bollo}. In this case the conductance
peaks are always symmetric, and their separation vs. temperature is shown in
inset (a). We also observe a sharp change in $G_N(T)$ at $T_N$: $G_N(T)$ is
even ten times larger for $T>T_N$ than for $T<T_N$ (inset (b)). This high
increase of $G_N$ would be in agreement with the fact that now both
electrodes have an increase of the electronic density of states for $T>T_N$.

We have repeated these experiments with UAu$_2$Si$_2$ in order to check
differences and similarities with URu$_2$Si$_2$. UAu$_2$Si$_2$ presents a
phase transition at $\sim 17$ K at zero pressure leading to a ferromagnetic
state, as it is observed in other compounds of this family, like UPt$_2$Si$%
_2 $. Conductance curves measured at 4.2 K present a single hump at zero
bias, symmetrical respect to voltage, which broadens and decreases its
height continuously, becoming almost negligible at 25 K (Fig.\ref{fig2}(B)).
In this case there is no sharp change in $G_N(T)$ during the thermal cycles.
This fact supports the previous discussion about the evolution of $G_N(T)$
in URu$_2$Si$_2$, because, to our knowledge, there is no evidence in the
literature about a loss of part of the Fermi surface in these heavy fermion
systems due to the phase transition.

We now interpret the resonances in URu$_2$Si$_2$ as arising from bound
states of conduction electrons and configurations of U$^{4+}$, in analogy to
the ``Kondo peak'' in spin 1/2 Kondo systems. We analyze the evolution of
the U$^{4+}$ configurations within the scheme proposed in\cite{mfield},
which is consistent with magnetic susceptibility and specific heat
experiments. A refinement of this analysis is presented in ref.\cite{spin}.
The scheme proposed in ref.\cite{mfield} analyses the low energy properties
of URu$_2$Si$_2$ in terms of the configurations of U$^{4+}$ only. It is
suggested in ref.\cite{spin} that the observed magnetization below $T_N$ in
this material can be due to an admixture of spin 1/2 configurations of U$%
^{3+}$ and U$^{5+}$ in a broken spin symmetry state. This mixing is assumed
to be small, in order to account for the small magnetic moment below the
transition temperature. In the following, we will interpret the present
experiments in terms of the low energy configurations of U$^{4+}$ only. We
do not expect significant changes due to the introduction of the U$^{3+}$
and U$^{5+}$ configurations. Note that, as far as the effect of the phase
transition on the configurations of U$^{4+}$ is concerned, refs.\cite{mfield}
and\cite{spin} are not contradictory.

According to ref.\cite{mfield}, the low energy properties of the U ions is
dominated by three low lying singlets. Hybridization with the conduction
band electrons are expected to induce transitions between these states, in a
similar manner to the quadrupolar Kondo fluctuations proposed for UBe$_{13}$%
\cite{qpolar}. In the present case, however, orthogonal symmetry breaks the
degeneracy of the two lowest lying singlets ($\Gamma _{t2}$ and $\Gamma
_{t1}^1$, in the notation of ref. \cite{mfield}).

The electron spectroscopy experiments show the development of a symmetric
resonance at zero voltage at low temperatures. We identify this resonance as
a Kondo-like peak\cite{resonance}. It arises from the scattering between the
two lowest lying U$^{4+}$ configurations, induced by the conduction
electrons. The associated Kondo temperature is $T_K\sim 60$ K. The fact that
the resonance is nearly symmetric around zero voltage implies that virtual
fluctuations into U$^{3+}$ and U$^{5+}$ charge states are equally likely.
The splitting between the low lying U$^{4+}$ configurations proposed in ref.%
\cite{mfield} is $\sim 9$ meV. As they are of the order, or less, than $T_K$%
, they cannot be resolved.

It is proposed in ref. \cite{mfield} that, below the transition, an
additional quadrupolar field acts on the U$^{4+}$ ions. This field mixes two
low lying configurations ($\Gamma _{t3}$ and $\Gamma _{t1}^1$), and
increases their separation in energies. In the presence of these effects,
the tunneling density of states should resemble that of the Kondo model in a
magnetic field\cite{Zeeman}, which shows two separate resonances. Within
this interpretation, the splitting between the two peaks that we find below $%
T_N$ should be directly proportional to the order parameter, and scale as $%
\sqrt{T_N-T}$. This behavior is, indeed, consistent with the results
reported here as shown in the good simulation presented in Fig.\ref{fig1}
(B), where the experimental curves have been modelled by two lorentzians
whose separation scales as $\sqrt{T_N-T}$, and different heights. The
inclusion of a third low energy resonance which is not changed by the
transition ($\Gamma _{t2}$) does not affect the quality of the result. An
alternative proposal, discussed in the literature, is the formation of an
antiferromagnetic gap at the Fermi surface, associated to the magnetic order
parameter. We find this interpretation less plausible than the one given
above as i) the AF order parameter is very small, and ii) it implies the
existence of a fully coherent heavy fermion band, with significant nesting
at the Fermi surface.

We finally address the question of the asymmetry between the two peaks which
arises at low temperatures. This asymmetry cannot be explained by the
opening of a gap at the Fermi level in an otherwise symmetric band. It does
not arise in a Kondo system in a magnetic field, where the two spin states
are split symmetrically. In the present case, a quadrupolar field mixes more
than the two configurations considered so far. The order parameter proposed
in ref.\cite{mfield}, $J_x^2-J_y^2$, hybridizes the two previously mentioned
states, $\Gamma _{t3}$ and $\Gamma _{t1}^1$ with $\Gamma _{t1}^2$, which
lies higher in energy. This additional hybridization also increases with the
order parameter, $\sim (T_N-T)^{1/2}$. The high energy configuration repels
the two low lying ones towards lower energies. The average shift in $\Gamma
_{t3}$ and $\Gamma _{t1}^1$ goes as the square of the order parameter, $\sim
T_N-T$. This displacement of the mean position is consistent with the
experiments, as also shown in Fig.\ref{fig1} (B).

Note, finally, that the splitting induced by a quadrupolar crystal field can
be much larger than typical spin splittings due to magnetic transitions. We
think that this is the reason for the unobservability of a splitting of the
Kondo resonance in UAu$_2$Si$_2$ below the ferromagnetic transition.

In summary, we have presented a series of electron spectroscopy experiments
using an STM, with the aim of investigating the 17.5 K phase transition in
URu$_2$Si$_2$. Our experimental conductance series, both for different
contact resistances at fixed temperature, and for different temperatures
keeping constant the resistance of the contact, suggest that the peaked
features in the spectra are consistent with the existence of ``Kondo like''
resonances built up from the different U$^{4+}$ configurations and the
conduction electrons. Below the transition temperature, these resonances are
split by the appearance of quadrupolar ordering. In addition, the ordering
induces an asymmetry in the density of states, associated to the large
number of ionic configurations affected by the transition.

We would like to thank R. Villar and C. Sirvent for their suggestions and
help, and D. Hinks, Argonne National Laboratory, Illinois, USA, for sample
preparation. This work has been supported by grants MAT94-0982, MAT95-1542,
and PB94-0382 (DGICyT).

\end{document}